\documentclass{article}
\usepackage{spconf,amsmath,graphicx}
\usepackage{tabularx}
\usepackage{commands}
\usepackage{amsmath}
\usepackage{amsfonts}
\usepackage{amssymb}
\usepackage{mathbbol}
\usepackage{color}
\usepackage{caption}
\usepackage{diagbox}
\usepackage{multirow}
\usepackage[T1]{fontenc}
\def\x{{\mathbf x}}

\title{Adaptive blind audio source extraction supervised by dominant speaker identification using x-vectors}
\name{Jakub Jansk\'{y}, Ji\v{r}\'{i} M\'{a}lek, Jaroslav \v{C}mejla, Tom\'{a}\v{s} Kounovsk\'{y}, Zbyn\v{e}k Koldovsk\'{y} and Jind\v{r}ich \v{Z}\v{d}\'{a}nsk\'{y}\thanks{This work was supported by the Technology Agency of the Czech
Republic (Project No. TH03010018) and by The Czech Science Foundation (Project No. 17-00902S).}}
\address{
Acoustic Signal Analysis and Processing Group\\ 
Faculty of Mechatronics, Informatics, and Interdisciplinary Studies, Technical University of Liberec\\ Studentsk\'a 2, 461 17
Liberec, Czech Republic}
\newcommand{\A}{{\bf A}_{k}}
\newcommand{\Ss}{{\bf S}_{k}}
\newcommand{\s}{{\bf s}_{k}}
\newcommand{\Aa}{{\bf A}_{k}}
\newcommand{\w}{{\bf w}_{k}}

\newcommand{\W}{{\bf W}_{k}}

\newcommand{\X}{{\bf X}_{k}}

\newcommand{\Pilot}{{\bf P}}
\newcommand{\PilotO}{{\bf P}_{\ell}^{\text{ORAC}}}
\newcommand{\PilotX}{{\bf P}_{\ell}^{\text{XVEC}}}
\newcommand{\splda}{ O({\bf s}_{\ell} )}
\newcommand{\zplda}[1]{ O({\bf y}^{#1}_{\ell})}
\newcommand{\shift}{{L}_{\text{shift}}}
\newcommand{\Lc}{L_{c}}

\newcommand{\Ak}{{\bf A}_{k}}

\newcommand{\zl}[1]{{\bf z}^{#1}_{k,\ell}}
\newcommand{\z}[1]{{\bf z}^{#1}_{k}}
\newcommand{\krok}{i}
\newcommand{\Vold}{{\bf V}_{k,i-1}}
\newcommand{\Vnew}{{\bf V}_{k,i}}
\newcommand{\wCxkold}{\hat{{\bf C}}_{k,i-1}}
\newcommand{\wCxknew}{\hat{{\bf C}}_{k,i}}

\begin{document}
\ninept
\maketitle
\begin{abstract}
We propose a novel algorithm for adaptive blind audio source extraction. The proposed method is based on independent vector analysis and utilizes the auxiliary function optimization to achieve high convergence speed.
The algorithm is partially supervised by a pilot signal related to the source of interest (SOI), which ensures that the method correctly extracts the utterance of the desired speaker. The pilot is based on the identification of a dominant speaker in the mixture using x-vectors. The properties of the x-vectors computed in the presence of cross-talk are experimentally analyzed. The proposed approach is verified in a scenario with a moving SOI, static interfering speaker and environmental noise.
\end{abstract}
\section{Introduction}
Independent Vector Analysis (IVA) performed in the frequency-domain is a popular approach to Blind Source Separation (BSS) of audio sources. It assumes that sources are mutually independent while separation proceeds jointly for all frequency bins by exploiting inter-channel higher-order dependencies. A variety of algorithms were proposed for IVA over the years, e.g., natural gradient~\cite{hiroe2006} or fast converging Auxiliary-function-based IVA (AuxIVA)~\cite{ono2011stable}. 

Typically, only a desired source should be extracted from the mixture in speech enhancement applications, which is the goal of Blind Source Extraction (BSE). A modification of IVA for the BSE problem, relating mixing and de-mixing vectors corresponding to the SOI through the orthogonal constraint (OG), was proposed in \cite{koldovsky2017b}.
In this work, a novel BSE algorithm referred to as AuxIVE (Auxiliary-function-based Independent Vector Extraction) is derived from AuxIVA by applying the OG. 
The method is adaptive in order to cope with possible movements and periods of inactivity of the SOI. A simmilar but non-adaptive modification of AuxIVA for BSE has been recently presented in \cite{scheibler2019independent}.

In BSE, there is the so-called global permutation problem that means that a different Interfering Source (IS) can be extracted instead of the SOI when no information about the SOI is available. In order to avoid this problem, a piloted IVA was introduced in \cite{pilot:intro} where the pilot signal related to the SOI is used for influencing the convergence to the SOI. However, the acquisition of a proper pilot signal poses a challenge.  
For example, a constrained location of the SOI has been exploited in \cite{pilot:intro}, additional audio measurements in the target area were used in \cite{pilot:bank}, or video information was used in \cite{pilot:video}. 

No additional measurements are required when the pilot is obtained through a pre-trained network. For example, a pilot based on voice activity detection was proposed in~\cite{pilot:video} for separating speech from non-speech. However, this solution fails when the interference is a concurrent utterance. In this work, we utilize a pilot that is based on the identification of a dominant speaker using speaker embeddings. This approach is applicable even when the observed signals contain non-speech noise as well as speech interference.

The concept of speaker embeddings has been introduced in the area of speaker recognition~\cite{xvector:speaker} and diarization~\cite{xvector:diary}. The goal is to map utterances to fixed-dimensional vectors which encode characteristics of the given speaker. Several embedding variants exist such as i-vectors~\cite{ivector} or the recently proposed Deep Neural Network-based (DNN) x-vectors~\cite{xvector:diary}. Very recently, the speaker embeddings have been utilized in a fully supervised speech extraction methods based on a pretrained dnn-based beamformer \cite{speakerbeam:intro,speakerbeam:journal}.
With a pretrained x-vector DNN, speaker identification is often performed using probabilistic linear discriminant analysis (PLDA, \cite{plda}). Here, a hypothesis is tested whether the embedding of an unknown speaker is produced by any of known speakers which are represented by embeddings computed from short clean utterances called enrollments.

It is usually assumed that the test signal contains a single speaker, nevertheless, multiple active speakers are possible as well \cite{xvector:multi}. 
In this work, we experimentally demonstrate that x-vectors can identify an active speaker even in the presence of cross-talk. This identification is reliable for the speaker with the highest energy in the mixture. 

The contribution of this paper is thus three-fold. First, the adaptive algorithm for piloted BSE is proposed. Second, speaker identification using x-vector~\cite{xvector:speaker} in the presence of cross-talk is investigated. 
Third, a pilot signal derived from the estimated dominance of the SOI is applied within the proposed algorithm. It is shown that short intervals of a dominant SOI activity are sufficient for the efficient dominance estimation and source extraction. We show that the method can be applicable also in situations where the SOI has lower global power than the other interfering sources. 







\section{Problem Description}
\label{sec:problem_description}

In the time-frequency domain, a mixture of $d$ original signals observed by $d$ microphones can be, within the $k$th frequency bin, approximated by the instantaneous mixing model
\begin{equation}
\X = \Aa \Ss
\label{eq:model}
\end{equation}
where $\Ss$ and $\X$ denote the original and mixed signals, respectively. In IVA, we are looking for a de-mixing matrix $\W$ that fulfills $\W\X = \W \Ak \Ss \approx \Ss$, which means $\W^{-1} \approx \A$, $k=1,\dots,K$.


\subsection{Independent Vector Extraction} 

In IVE, only one row of $\W$ is sought such that extracts the SOI from the mixture. Without any loss on generality, let the SOI be the first signal in $\Ss$ and $\A$ be partitioned as $\A=[{\bf a}_k, {\bf Q}_k]$. Then, \eqref{eq:model} can be expressed in the form
\begin{equation}
   \begin{pmatrix}
x^{1}_{k,1} & \cdots & x^{1}_{k,L} \\
x^{2}_{k,1} & \cdots & x^{2}_{k,L} \\
\vdots & \ddots &\vdots\\
x^{d}_{k,1} & \cdots & x^{d}_{k,L} 
\end{pmatrix} = [{\bf a}_k, {\bf Q}_k]\begin{pmatrix} s_{k,1} & \cdots & s_{k,L} \\
z^{2}_{k,1} & \cdots & z^{2}_{k,L} \\
\vdots & \ddots &\vdots\\
z^{d}_{k,L} & \cdots & z^{d}_{k,L} 
\end{pmatrix},  
\end{equation}
where $k = 1,\dots,K$, $\ell = 1,\dots,L$ is the frame index, $\s = [s_{k,1}, \cdots , s_{k,L}]$ represents the SOI, $\z{j}$ for $j = 2,\dots,d$ are the other original signals in the mixture, and $\xkl = [x^{1}_{k,\ell},\dots,x^{d}_{k,\ell}]^T$ corresponds to the signals on microphones. Using the parameterization from  \cite{koldovsky2019TSP}, $\W$ can be partitioned as $\W=[({\bf w}_k)^H;\, {\bf B}_k]$, where $\w$ is a vector extracting the SOI, and ${\bf B}_k$ is a sub-matrix whose rows are orthogonal to ${\bf a}_k$.


Now, we apply the IVE statistical model from \cite{koldovsky2019TSP}, so the contrast function (for one signal sample) for the estimation of parameter vectors ${\bf a}_k$ and ${\bf w}_k$, $k=1,\dots,K$, is given by
\begin{multline}\label{eq:contastIVE}
\mathcal{J}(\{{\bf w}_k\}_{k=1}^K,\{{\bf a}_k\}_{k=1}^K)=\log 
f\bigl(\widehat{\bf s}_1,\dots,\widehat{ \bf s}_K\bigr)  \\
-\frac{1}{L}\sum_{k=1}^{K}\sum_{\ell=1}^{L}\xkl^H{\bf B}_k^H{\bf C}_{{\bf z}_k}^{-1}{\bf B}_k\xkl +\sum_{k=1}^K\log|\det{\bf W}_k|^2,
\end{multline}
where $\widehat{s}_{k,\ell}=({\bf w}_k)^H\x_{k,\ell}$, $f(\cdot)$ is the model probability density function of SOI and ${\bf C}_{{\bf z}_k}$ is the covariance of the interference signals (assumed to be Gaussian). Furthermore, to keep $\w$ and ${\bf a}_k$ more tightly connected, the OG is imposed through
\begin{equation}\label{eq:OG}
    {\bf a}_{k} =\frac{\wCxk{\bf w}_k}{{\bf w}_k^H\wCxk{\bf w}_k},
\end{equation}
where $\wCxk$ is the sample covariance matrix of the mixture.

\subsection{Adaptive AuxIVE Algorithm}

For the optimization of the contrast function given by \eqref{eq:contastIVE}, we can apply the auxiliary function technique in a similar way to \cite{ono2011stable}. 
By stating the same assumption about the model density $f(\cdot)$ as in Theorem~1 in \cite{ono2011stable}, the auxiliary function for \eqref{eq:contastIVE} (for all available samples) can have the form
\begin{multline}\label{eq:auxiliaryfunctionIVE}
    \mathcal{Q}(\{{\bf w}_k\}_{k=1}^K,\{{\bf a}\}_{k=1}^K,r_{\ell})= -\frac{1}{2}\sum_{k=1}^K({\bf w}_k)^H {\bf V}_k{\bf w}_k \\-\frac{1}{L}\sum_{k=1}^{K} \sum_{\ell=1}^{L}\xkl^H{\bf B}_k^H{\bf C}_{{\bf z}_k}^{-1}{\bf B}_k\xkl +\sum_{k=1}^{K}\log |\det{\bf W}_k|^2 + R,   
\end{multline}
where 
\begin{equation}\label{eq:Vk}
    {\bf V}_k = \frac{1}{L}\sum_{\ell = 1}^L\varphi(r_{\ell})\xkl\xkl^H,
\end{equation}
and $r$ is the auxiliary variable, which is scalar in this case; $R$ is a constant term, and $\varphi(\cdot)$ is related to the $f(\cdot)$ according to Theorem 1~\cite{ono2011stable}. The equality between \eqref{eq:contastIVE} and \eqref{eq:auxiliaryfunctionIVE} holds if and only if $r_{\ell}=\sqrt{\sum_{k=1}^K |({\bf w}_k)^H\xkl|^2}$. 

The derivative of \eqref{eq:auxiliaryfunctionIVE} with respect to ${\bf w}_k^H$ under the OG \eqref{eq:OG} has the closed solution form 
\begin{equation}
    \w =  {\bf V}_k^{-1}{\bf a}_k.
\end{equation}

Considering the sequential (block-by-block) adaptive processing of data blocks with a length of $L_b$ frames with shift $\shift$, the update rules for the $\krok$th block have a form
\begin{align}
    r_{\ell,i}&=\sqrt{\sum\nolimits_{k=1}^K |{\bf w}_{k,\krok-1}^H \xkl|^2} \qquad \text{for } \ell = \ell_{s},\dots, \ell_{e} \label{eq:update_rule}\\
    \Vnew &= \alpha\Vold  + (1-\alpha)\frac{1}{L_b}\sum\nolimits_{\ell = \ell_{s}}^{\ell_{e} }[\varphi(r_{\ell})\xkl \xkl^H],\\
    \wCxknew &= \alpha\wCxkold + (1-\alpha) \frac{1}{L_b}\sum\nolimits_{\ell = \ell_{s}}^{\ell_{e} } \xkl \xkl^H\\
    {\bf a}_{k,\krok} &=\frac{\wCxknew{\bf w}_{k,\krok-1}}{{\bf w}_{k,\krok-1}^H\wCxknew{\bf w}_{k,\krok-1}},\\
  {\bf w}_{k,\krok} & =  \Vnew^{-1}{\bf a}_{k,\krok},\label{eq:step5}
\end{align}
where $\alpha$ is a forgetting factor; $\ell_{s} = (\krok-1)\shift +1$ and $\ell_{e} = (\krok-1)\shift +L_b$ denote the beginning and the end of the $\krok$th block, respectively.
For the case $L_b = 1$ and $\alpha \in (0,1\rangle$, we refer to the proposed method as to ``Online AuxIVE'', and, for $L_b > 1$ and $\alpha = 0$, we call it ``Block Online AuxIVE''. Note that, for $L_b = 1$, the matrix inversion lemma can be applied for the fast computation of \eqref{eq:step5}.


\subsection{Supervised AuxIVE}
To ensure the extraction of the desired source, we propose to employ a pilot component in the same manner as in \cite{nesta2017supervised}. Let $\Pilot$ be a pilot signal that is SOI-dependent and independent with  the other signals in the mixture. $\Pilot$ is independent of the mixing model parameters and thus does not change the analytic learning rules of AuxIVE up to the difference that the non-linearity $\varphi(\cdot)$ depends on $\Pilot$. Consequently, the step given by \eqref{eq:update_rule}, for the $i$th block, is modified to
\begin{equation}\label{eq:auxivepiloted}
r_{\ell,i}=\sqrt{\sum\nolimits_{k=1}^K |{\bf w}_{k,\krok-1}^H \xkl|^2 + \Pilot_{\ell}}, \quad  \ell = \ell_{s},\dots, \ell_{e}.
\end{equation}  
The rest of the algorithm remains the same.

\section{X-vector computation}
\label{sec:xvector}

\subsection{The x-vector DNN}

Our implementation of the x-vector DNN, described in Table~\ref{tab:xvecnet}, stems from paper~\cite{xvector:speaker}. Its input consists of a single-channel audio signal, it does not use any spatial information. The input features are $40$ filter bank coefficients computed from frames of length of $25$~ms and frame-shift of $10$~ms. The TDNN (time-delayed DNN) layers introduced in~\cite{tdnn:intro} operate on frames with a temporal context centered on the current frame $\ell$. The TDNN layers build on top of the context of the earlier layers, thus the final context is a sum of the partial ones. In contrast to~\cite{xvector:speaker}, we introduced four differences in the DNN: 1) We use all the frames in the context without any sub-sampling, in order to exploit the time-dependencies in the signal. 2) At the input of each TDNN layer, we weight all the frames in the context by a trainable matrix and perform mean time-pooling, in order to limit the number of trainable parameters. 3) We replaced the rectified linear units at the output of TDNN and fully-connected layers by exponential linear units (ELU), which speeds up convergence in our case. 4) The pooling layer computes only variances of frames (means are omitted) in the context, which is during the training phase set to $\Lc=151$.

The DNN was trained to classify $N$ speakers. The training examples consisted of $151$ frames of features and the speaker label. The data originate in the development part of the Voxceleb database~\cite{voxceleb}. 

\begin{table}[]
\begin{center}
\caption{\label{tab:xvecnet} Description of the DNN producing the x-vectors. The input size for the TDNN layers is stated after the mean pooling operation.}
\begin{tabular}{|c|c|c|c|}
 \hline
 \textbf{Layer} & \textbf{Layer} & \textbf{Total} & \textbf{Inpup} \\
  & \textbf{context} & \textbf{context} & x \textbf{output} \\ 
 \hline\hline
TDNN 1  & $\ell\pm50$ & $101$ & $40 \times 512$ \\ \hline
TDNN 2-6  & $\ell\pm5$ & $151$ & $512 \times 512$ \\ \hline
Fully-conn. 1 & $\ell$ & $151$ & $512 \times 128$ \\ \hline
Pooling & $\ell\pm \frac{\Lc-1}{2}$ & max$(151,\Lc)$ & $(\Lc\cdot128)\times 128$ \\ \hline
Fully-conn. 2  & $\ell$ & max$(151,\Lc)$ & $128\times 128$ \\ \hline
Softmax  & $-$ & max$(151,\Lc)$ & $128\times N$ \\ \hline
\end{tabular}
\end{center}
\end{table}

\subsection{Speaker identification in the presence of cross-talk}

During test phase, the x-vectors are extracted at the output of the pooling layer. In order to obtain time-localized x-vectors, the context of the DNN is shifted by a single frame through the utterance.
The assignment of test recording to the enrollment speakers is performed via PLDA~\cite{plda}, trained using the test part of the Voxceleb database~\cite{voxceleb}. Prior the PLDA modelling, no dimensionality reduction is performed, the vectors are centered and length-normalized. 

For the dominant speaker identification, we assume a scenario with a closed set of a small number of known speakers. We show in Section~\ref{sec:exp:activity} that if the test signal contains two active speakers from such a limited enrollment set, the PLDA score is usually highest for the dominant speaker from the perspective of signal energy. However, the second active speaker cannot be reliably determined in this manner, the PLDA score often suggests a non-active speaker. 

\subsection{Pilot signal estimation}

The pilot signal needs to be dependent on SOI. It should exhibit high values when SOI is active and vice versa. Thus, we define the ``oracle" energy-based pilot $\PilotO$ in the $\ell$th frame as
\begin{equation}\label{eq:oracle_pilot}
\PilotO =\begin{cases}
\sum_{k=1}^K|\Xkl|^2 & \frac{\sum_{k=1}^K|\skl|^2}{\sum_{j = 2}^d\sum_{k=1}^K|\zl{j}|^2}  \geq \nu,\\
0 & \text{otherwise},\\
\end{cases}
\end{equation}
where $\nu$ is a free threshold parameter and $\zl{j}$ for $j = 2,\dots,d$ is the $j$th original signal not related to SOI. Signal $\PilotO$ cannot be computed in practice, as it depends on the unknown energies of the underlying sources/utterances. It serves us as an ideal example of the functionality of the pilot.

Based on observations from Section~\ref{sec:exp:activity}, the highest PLDA score corresponding to SOI is usually related to the frames where SOI is dominant. In a manner similar to~\eqref{eq:oracle_pilot}, we propose the pilot $\PilotX$ as
\begin{equation}\label{eq:PLDA_pilot}
    \PilotX =\begin{cases}
    \sum_{k=1}^K|\Xkl|^2 & \frac{\splda}{\max(\zplda{1},\dots,\zplda{M})}  \geq \eta,\\
    0& \text{otherwise,}\\
    \end{cases}
\end{equation}
where $\splda$ is the PLDA score of SOI, $\zplda{m}$ for $m = 1,\dots,M$ is the PLDA score of the enrollment speakers other than SOI and $\eta$ is a free threshold parameter.


\section{Experimental evaluation}
\label{exp:room_setup}
\begin{figure*}
  \begin{minipage}{.2\textwidth}
    \includegraphics[width=\linewidth]{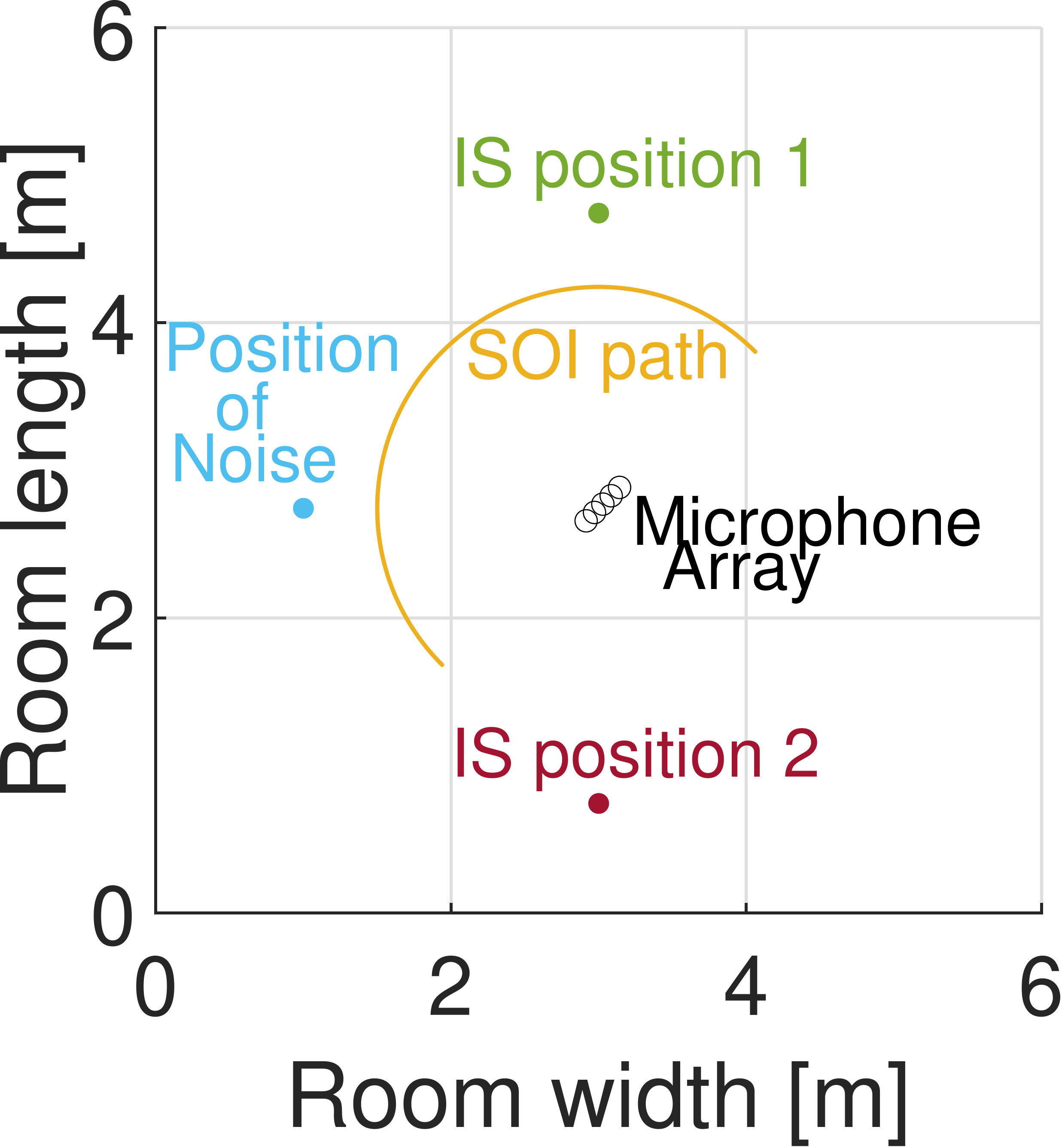}
    \caption{\label{fig:room_setup} Setup of the simulated room scenario}
    \label{tab:results}
  \end{minipage} \quad
  \begin{minipage}{.70\textwidth}
    \captionof{table}{Mean iSNR [dB] with standard deviation and percentage of failed extractions ($\text{iSNR}<1$) for experiment in~Section~\ref{sec:exp:exp3}. The mixtures with IS in position 1 are prone to extraction of an undesired source.}
    \setlength\tabcolsep{5pt}
    \begin{tabular}{|c|r|ccc|ccc|}
        \cline{3-8} 
        \multicolumn{2}{l|}{} & \multicolumn{3}{c|}{Block online AuxIVE} & \multicolumn{3}{c|}{Online AuxIVE} \\ \cline{3-8} 
        \multicolumn{2}{l|}{} & 
        \rotatebox[origin=c]{47}{Blind} &
        \rotatebox[origin=c]{47}{$\PilotX$} &
        \rotatebox[origin=c]{47}{$\PilotO$} &
        \rotatebox[origin=c]{47}{Blind} &
        \rotatebox[origin=c]{47}{$\PilotX$} &
        \rotatebox[origin=c]{50}{$\PilotO$}  \\ \hline
        \multirow{2}{*}{\parbox{1.6cm}{IS position 1}} & iSNR {[}dB{]} & 4.3 $\pm$ 3.6 & 6.4 $\pm$ 1.9  & 10.0 $\pm$ 1.7 & -0.5 $\pm$ 1.8 & 2.0 $\pm$ 1.5  & 5.0 $\pm$ 1.5  \\
        & fail cases [\%]& 24.67 & 2 & 0 & 77.67 & 23.67 & 1.67 \\ \hline
        \multirow{2}{*}{\parbox{1.6cm}{IS position 2}} & iSNR {[}dB{]} & 9.3 $\pm$ 1.7  & 9.6 $\pm$ 1.5  & 12.6 $\pm$ 1.8  & 5.0 $\pm$ 1.6  & 5.9 $\pm$ 1.4  & 8.5 $\pm$ 1.5  \\
        & fail cases [\%] & 0 & 0 & 0 & 0.34 & 0 & 0 \\\hline
        \multicolumn{2}{|c|}{Average time per mixture{[}s{]}} & 4.55 & 11.8 & 4.65 & 15.24 & 24.42 & 15.32 \\ \hline
    \end{tabular}
  \end{minipage}
  \vspace*{-0.7cm}
\end{figure*}

The experiments simulate speaker movements in a reverberant room of dimension $6 \times 6 \times 3$ m, with $T_{60} = 100$~ms. An array of five omni-directional microphones was placed close to the center of the room and rotated counter-clockwise by $45^{\circ}$. SOI was moving $40$~cm/s in a semicircle 1.5 m in front of the array. The IS was placed either at coordinates $(3, 4.74)$ or $(3,0.74)$, see Fig.~\ref{fig:room_setup}. The distance from both IS positions to the microphone array was 2 m. 

We utilized signals originating from the CHiME-4 database, namely the simulated part of the test and development sets. We concatenated one minute of single-channel noiseless speech from four speakers (F01, F06, M04, M05) for the enrollment.
For the test signals, another utterances of each considered speaker, distinct to the enrollment, were concatenated. We obtained $5$ unique one minute long test signals sampled at $16$~kHz. The RIR generator \cite{habets2006room} was used to simulate the source movements. Noise signal, positioned at $(1,2.74)$, consists of one minute of pedestrian area sounds.

The experiments are evaluated using improvement of Signal-To-Noise ratio (iSNR), where all undesired sources are included in the noise term. We also provide attenuation of SOI defined as $10\log_{10} \sum_k |\hat{s}_{k,\ell}|^2 / \sum_k|s_{k,\ell}|^2$. Ideally, the extracted source should have constant attenuation with respect to time index $\ell$, i.e., exhibit no loudness fluctuations.

\renewcommand{\arraystretch}{1}

Free parameters of the proposed methods were set in the following manner. For Block online AuxIVE, $L_b =100$, $\shift = 75$ and $\alpha = 0$. For Online AuxIVE, $L_b =1$, $\shift = 1$ and $\alpha = 0.97$. Both methods used $K=512$ with frame shift $160$ samples (that is $10$~ms as in x-vector DNN). The demixing vector updates were computed using single iteration in every block. Oracle pilot~\eqref{eq:oracle_pilot} used $\nu= 0.5$; x-vector-based pilot~\eqref{eq:PLDA_pilot} used $\eta = \exp(-5)$. 

\subsection{Case study: dominant speaker identification}
\label{sec:exp:activity}

This sections shows the basic properties of x-vectors, when computed on a test signal with overlapping utterances. We analyze one minute of mixed noiseless speech originating from speakers F01 and M04; global energy of F01 being slightly higher by $2$~dB.

\begin{figure}
    \begin{center}
	\includegraphics[width=0.49\linewidth]{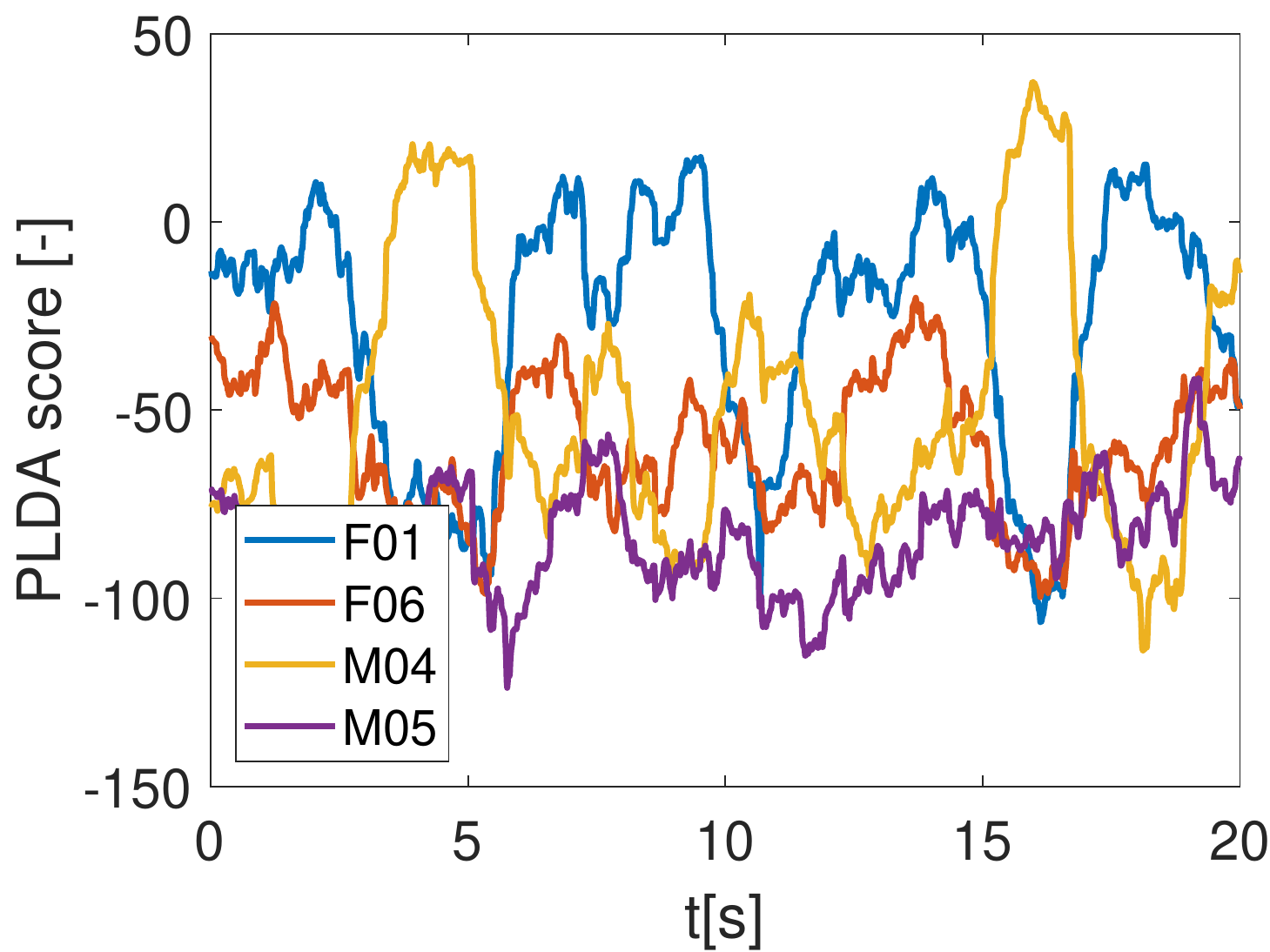}
	\includegraphics[width=0.49\linewidth]{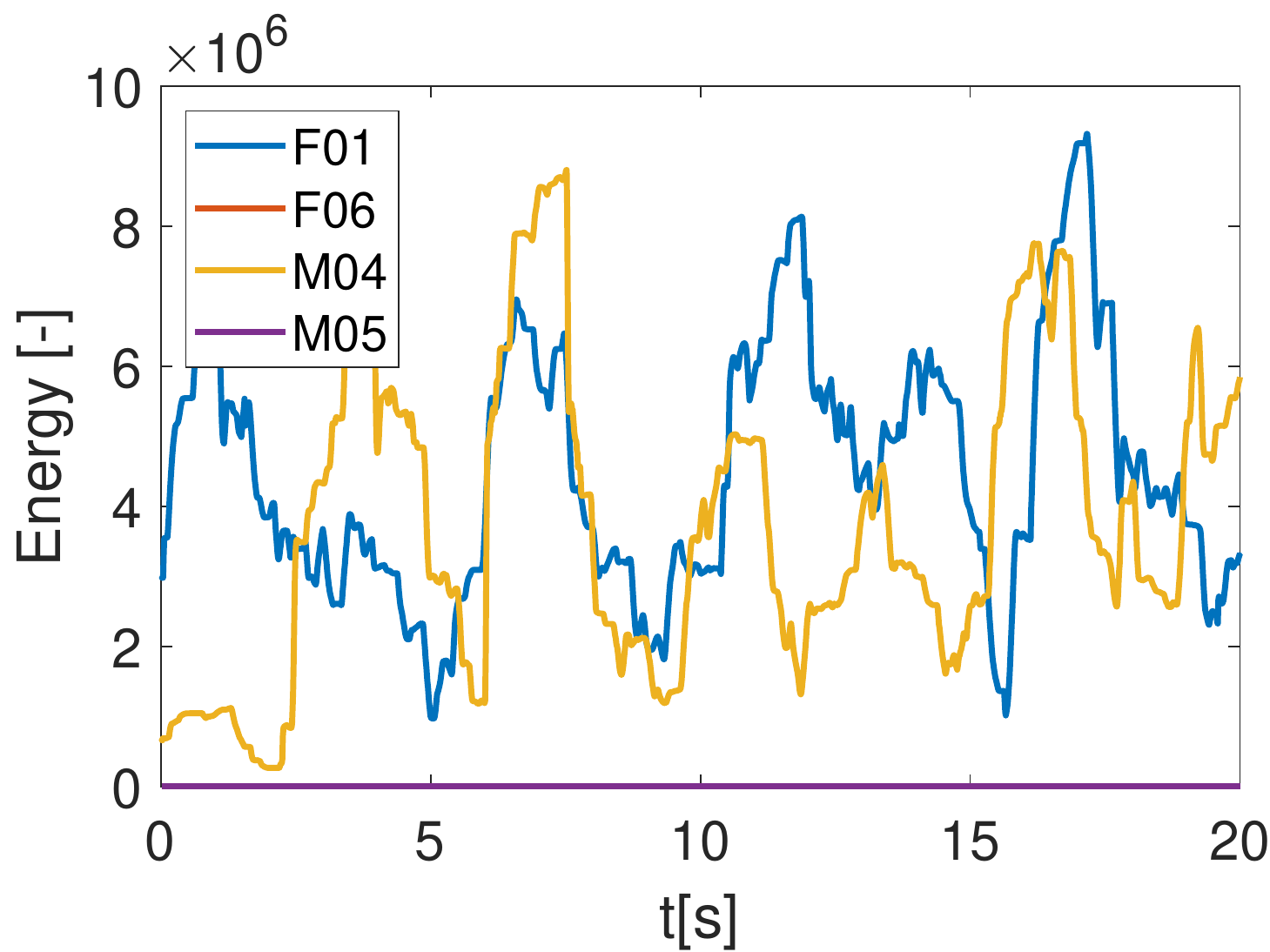}
	\caption{\label{fig:case-study} PLDA score and the corresponding energy of the respective utterances for the mixture discussed in Section~\ref{sec:exp:activity}.}
    \end{center}
    \vspace*{-0.7cm}
\end{figure}

The Figure~\ref{fig:case-study} shows for the first twenty seconds of the mixture (from the first microphone), the obtained PLDA score $O(\cdot)$ and the signal energies of the underlying utterances. The energy is computed within the same context centered around the current frame as utilized by the DNN (in this case $T=151$, i.e., $1.5$~s).
It can be seen that the highest PLDA score corresponds to the speaker with the highest energy most of the time. However, the second highest PLDA score often does not correspond to the second active speaker. For example, in interval $0-3$~s the second speaker according to PLDA score is F06, which is not active in the recording. For the whole mixture, the dominant speaker is determined with $79.8\%$ accuracy. Only in $0.1\%$ cases, one of the non-active speakers (F06, M05) is estimated as dominant.
 
To construct the pilot $\PilotX$ via~\eqref{eq:PLDA_pilot}, we shorten the context of the pooling layer within the x-vector DNN to $\Lc=10$ frames, in order to obtain more time-localized dominance estimate. However, usage of the shorter context lowers the accuracy of the speaker-dominance estimate to $62.4\%$. Also, the non-active speaker is miss-classified as dominant more often, in $21.7\%$ cases. However, to construct $\PilotX$ we actually classify a simpler problem: we want do distinguish dominant/non-dominant frames for SOI only, not all of the potential speakers. If F01 is the SOI, this accuracy is $68.9\%$; for M04 the accuracy is $77.7\%$.

\subsection{Case study: SOI extraction in cross-talk}
\label{sec:exp:exp2}
This section extends the case study discussed in Section~\ref{sec:exp:activity}. We create a mixture of moving speaker M04 as SOI and speaker F01 from position 1 as IS. Input SNR of the mixture was -2.68 dB. We compare SOI extracted by the Block AuxIVE with and without pilot.

\begin{figure}[h!]
    \begin{center}
	\includegraphics[width=0.49\linewidth]{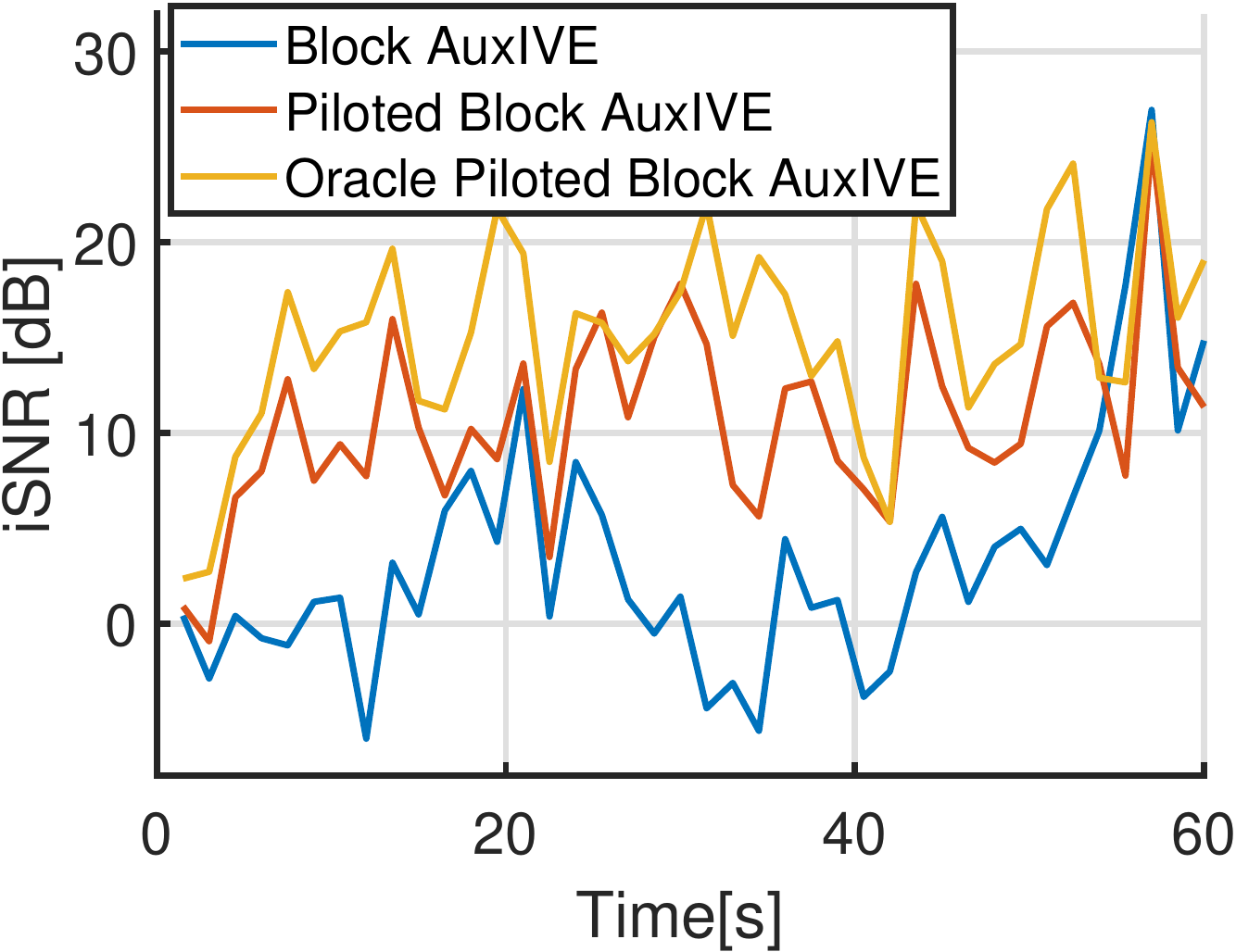}
	\includegraphics[width=0.49\linewidth]{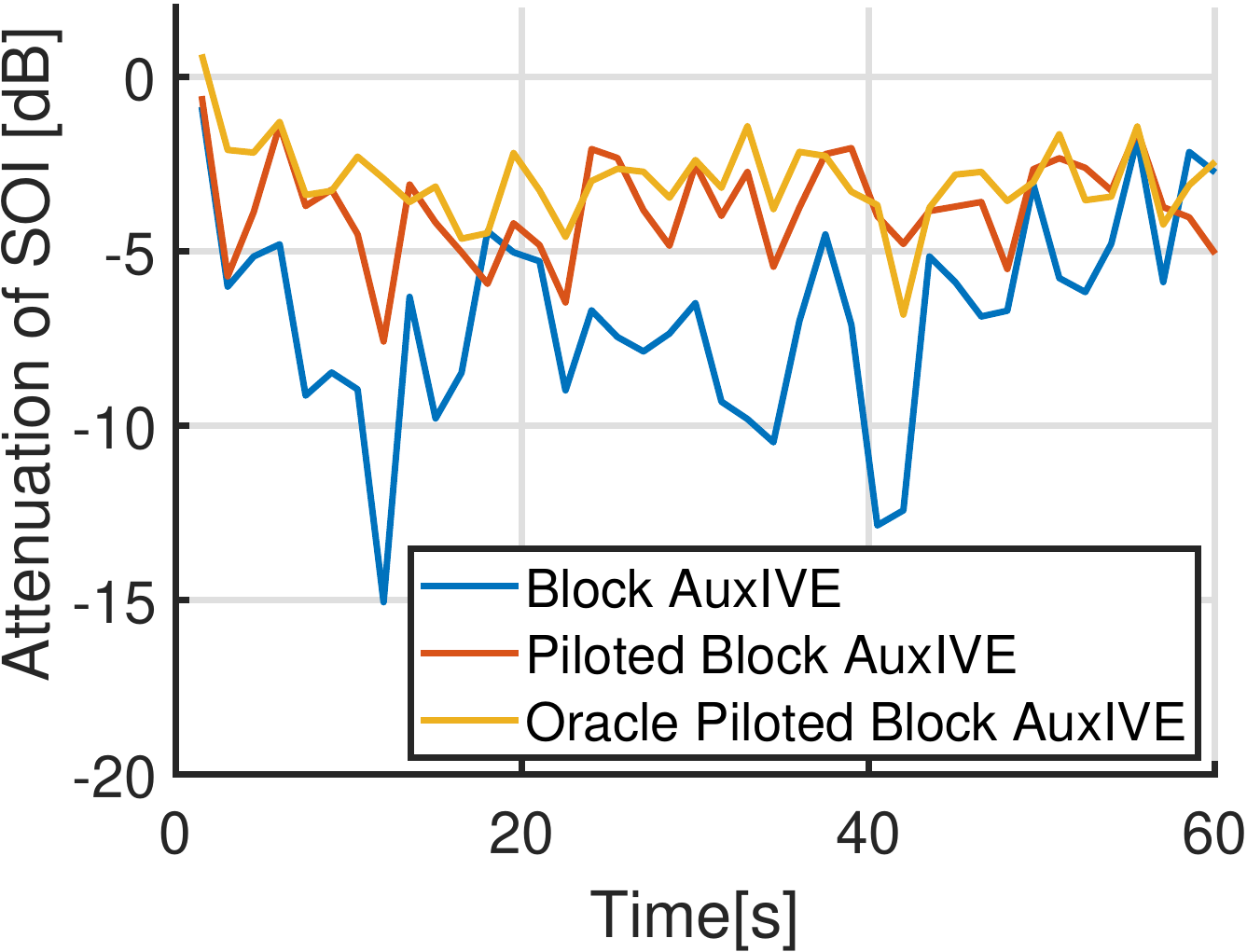}
	\caption{Time dependence of iSNR and SOI attenuation for case study discussed in Section~\ref{sec:exp:exp2}.}
	\label{fig:epx2_results}
    \end{center}
    \vspace*{-0.7cm}
\end{figure}

The average iSNR was only 1.66~dB for the blind Block AuxIVE. The dependency of iSNR on time in Figure~\ref{fig:epx2_results} suggests
that the blind version suffers with the global permutation problem, as the iSNR drops bellow zero several times, i.e., the incorrect source was extracted. The utilization of $\PilotX$ increases the average iSNR to $7.12$~dB and no source permutation is observed.

The Figure~\ref{fig:epx2_results} shows that blind AuxIVE does have very variable SOI attenuation, because it changes its focus on and off the SOI. The piloted version using $\PilotX$ achieves more stable one, approaching utilization of the oracle $\PilotO$.

\subsection{Extraction in noisy environment}
\label{sec:exp:exp3}

This section discusses average results yielded by the proposed methods extracting SOI from $600$ simulated mixtures ($6$ speaker combinations $\times 2$ speaker roles $\times 25$ utterance combinations $\times 2$ IS positions). 
Each mixture consists of a moving SOI, a fixed IS and noise (with global energy $10$~dB lower compared to the summed utterances). The IS was situated either in position 1, where the methods are prone to permutation problem due to close proximity to SOI path or position 2, which should not have such problem. The input SNR was approximately the same for all mixtures at 1.35~dB on average. We consider the SOI extraction as failed, when $\text{iSNR} < 1$. 

The results in Table~\ref{tab:results} indicate
that the blind variants of AuxIVE yield positive iSNR for most of the considered mixtures. The introduction of a pilot improves the extraction especially for cases with IS in position 1. The higher iSNR along with its lower variance and lower fail-rate show that the utilization of pilot successfully prevents the source permutation.

In comparison to $\PilotO$, the supervision by $\PilotX$ brings lower improvements in average iSNR. However, it never leads to major deterioration compared to the blind method. Considering IS in position 1, $\PilotX$ improves the iSNR in $83\%$ of cases for Block online AuxIVE and $100\%$ cases for the online version. The worst case deterioration due to use of $\PilotX$ is always lower than $1$~dB.

\section{Conclusion}

A blind adaptive and fast converging method for BSE was proposed, suitable for SOI extraction in the noisy cross-talk scenario.
To avoid extraction of an incorrect source, a pilot related to SOI activity was presented based on x-vectors. For environments with low reverberation, it was shown that the x-vectors can identify the energy-dominant speaker even in the presence of cross-talk.

The future research can be directed towards improvement of the pilot. The x-vector DNN can be adapted to function even for higher reverberation levels (e.g. through augmentation of the training data) or 
to produce more time-localized estimates of the active speaker.



\end{document}